\documentclass{article}

\usepackage{arxiv}
\usepackage[utf8]{inputenc} 
\usepackage[T1]{fontenc}    
\usepackage{hyperref}       
\usepackage{url}            
\usepackage{booktabs}       
\usepackage{amsfonts}       
\usepackage{nicefrac}       
\usepackage{microtype}      
\usepackage{lipsum}		
\usepackage{amsmath}
\usepackage{graphicx}
\usepackage{natbib}
\usepackage{doi}
\usepackage{graphicx}
\usepackage{bm}
\usepackage{amssymb}
\usepackage{multicol}
\usepackage{multirow}
\usepackage{float}
\usepackage{enumitem}
\usepackage[capitalize]{cleveref}
\usepackage[title]{appendix}
\usepackage{icomma}
\usepackage{caption}
\usepackage{subcaption}

\newcommand{\E}{\text{E}}
\newcommand{\Var}{\text{Var}}

\newcommand{\dirch}{\text{Dirichlet}}
\newcommand{\mulnom}{\text{Multinom}}

\newcommand{\iid}{\overset{iid}{\sim}}
\newcommand{\mmu}{\bm{\hat{p}}}
\newcommand{\pp}{\bm{\pi}}
\newcommand{\XX}{\bm{X}}
\newcommand{\rhotk}{\rho_{k,t}}
\newcommand{\ptk}{\pi_{k,t}}
\newcommand{\mut}{\hat{p}_t}
\newcommand{\mutk}{\hat{p}_{k,t}}
\newcommand{\muk}{\hat{p}_{k}}
\newcommand{\sgtk}{s_{k,t}}
\newcommand{\sgk}{s_{k}}
\newcommand{\muobs}{\hat{p}_{k,t}^{\text{obs}}}
\newcommand{\muobsk}{\hat{p}_{k}^{\text{obs}}}
\newcommand{\sgobs}{s_{k,t}^{\text{obs}}}
\newcommand{\sgobsk}{s_{k}^{\text{obs}}}
\newcommand{\w}{w_t}
\newcommand{\bw}{\bm{w}}
\newcommand{\bs}{\bm{s}}
\newcommand{\n}{n_t}
\newcommand{\Nhat}{\hat{N}}
\newcommand{\lam}{\lambda_t}
\newcommand{\lamt}{\tilde{\lambda}_t}
\newcommand{\betat}{\tilde{\beta}_t}
\newcommand{\sgalt}{\hat{\sigma}_k^\text{2,alt}}

\title{Rapid Scaling of Compositional Uncertainty from Sample to Population Levels}

\date{} 					

\author{Yiran Wang \quad  \quad  \\
	Department of Biostatistics\\
	Yale School of Public Health\\
	New Haven, Connecticut, U.S.A 06510\\
	\And
	Martin Lysy \\
	Department of Statistics and Actuarial Science\\
	University of Waterloo\\
	Waterloo, Ontario, Canada N2L 3G1 \\
	\AND
	Audrey B\'eliveau$^*$ \\
	Department of Statistics and Actuarial Science\\
	University of Waterloo\\
	Waterloo, Ontario, Canada N2L 3G1 \\
        \texttt{audrey.beliveau@uwaterloo.ca}\\
}

\renewcommand{\headeright}{A preprint}
\renewcommand{\undertitle}{A preprint}
\renewcommand{\shorttitle}{Rapid Scaling of Compositional Uncertainty from Sample to Population Levels}

\begin{document}
\maketitle

\begin{abstract}
Understanding population composition is essential across ecological, evolutionary, conservation, and resource management contexts. Modern methods such as genetic stock identification (GSI) estimate the proportion of individuals from each subpopulation using genetic data. Ideally, these estimates are obtained through mixture analysis, which captures both sampling and genetic uncertainty. However, historical datasets often rely on individual assignment methods that only account for sample-level uncertainty, limiting the validity of population-level inferences. To address this, we propose a reverse Dirichlet-multinomial model and derive multiple variance estimators to propagate uncertainty from the sample to the population level. We extend this framework to genetic mark-recapture studies, assess performance via simulation, and apply our method to estimate the escapement of Sockeye Salmon (\textit{Oncorhynchus nerka}) in the Taku River.
\end{abstract}


\section{Introduction} \label{taku_intro}

Estimating the composition of a population---i.e., the proportion of individuals belonging to each of several subpopulations---is a central task in ecological, evolutionary, conservation, and management settings~\citep[e.g., ][]{allendorf2012conservation}. Let $\pp = (\pi_1,\ldots,\pi_K)'$ denote the vector of population proportions, where each $\pi_k$ represents the proportion of individuals from subpopulation $k$, and $\sum_k\pi_k=1$. Suppose we obtain a simple random sample of $n$ individuals from the population. If the true subpopulation label for each sampled individual were known, then the number of individuals from each group, denoted by $\XX = (X_1,\ldots,X_K)'$, would follow a multinomial distribution: $\XX\sim \mulnom(n,\pp)$. In this ideal case, estimation of $\pp$ and its uncertainty would follow directly.

However, in many real-world applications, including those involving genetic stock identification (GSI), the subpopulation label for each sampled individual is unknown. Instead, probabilistic methods are used to infer subpopulation membership based on genetic markers, comparing observed genotypes with those from baseline reference populations~\citep{grant1980use,Pella2001Bayesian,hess2014monitoring}. A principled approach is mixture analysis (MA), which estimates $\pp$ by jointly modeling population sampling and genetic classification, thereby yielding composition estimates with uncertainty measures that reflect both sources of variability~\citep[e.g.,][]{koljonen2005classical,kuismin2020genetic}. Yet in practice---particularly in historical datasets---analyses are often based on individual assignment (IA), where each sampled individual is assigned a set of membership probabilities, and population-level composition is obtained by averaging these across individuals. When subpopulations are highly separable genetically, these individual probabilities collapse toward 0 or 1, so the averaged proportions are essentially the observed sample fractions with little reported uncertainty. As a result, IA-based uncertainty estimates typically neglect sampling variation and underestimate the true variability in $\pp$, limiting their use for population-level inference.

To address this challenge, we develop post-hoc strategies that correct IA-based uncertainty estimates to reflect sampling variability at the population level. Crucially, our approach treats the IA procedure as a black box, requiring only the reported point estimates and uncertainty summaries for each $\pi_k$. This enables the propagation of uncertainty to the population level without reanalyzing the original genetic data or replicating the IA procedure.

Our core proposal is a reverse Dirichlet-multinomial (RDM) model, which treats the IA estimates as realizations from a Dirichlet distribution centered on the true (multinomial) sample composition. This allows us to link sample-level estimates to population-level uncertainty through a flexible generative model. We also introduce a moment-matching Dirichlet (MMD) model as a computationally efficient alternative to the RDM, obtained by matching the first two moments of the IA estimates without explicitly modeling latent counts. Both models yield closed-form expressions for population-level variance, enabling the construction of multiple variance estimators for $\pp$.

We further extend our framework to genetic mark-recapture (GMR), a method widely used in mixed-stock salmon fisheries~\citep{hamazaki2014application}. In this context, escapement refers to the number of adult salmon that migrate from the ocean back to their natal freshwater habitats to spawn, avoiding harvest along the way. Estimating escapement by stock is central to fisheries management, as it enables managers to set harvest targets that reflect the relative abundance of different stocks~\citep{ostergren2020close}. GMR is particularly valuable when escapement counts are nearly complete for some stocks---for example, lake-type stocks that can be enumerated at counting weirs---while other stocks, such as river-type populations, must be identified in genetic samples collected during the migration. In this setting, escapement is estimated by dividing the escapement count of lake-type stocks by their proportion in the GSI sample, producing a Petersen-type estimator. However, when the GSI proportions come from IA-based analyses, the reported uncertainty typically reflects only sample-level variation. Our framework propagates this uncertainty to the population level, yielding valid escapement uncertainty estimates. To further improve inference, we develop a Bayesian version of our method that incorporates temporal autocorrelation in weekly composition through a novel autoregressive prior.

The remainder of this paper is structured as follows. Section~\ref{taku_method} introduces two complementary models for adjusting IA-based composition estimates: the reverse Dirichlet-multinomial (RDM) model and the moment-matching Dirichlet (MMD) model, and describes how both yield closed-form variance estimators for population composition. Section~\ref{taku_gmr} applies these tools to GMR, presenting both Bayesian and frequentist formulations for escapement estimation. Section~\ref{taku_sim} presents simulation results evaluating model performance. Section~\ref{taku_application} applies our methods to estimate the escapement of Sockeye Salmon (\textit{Oncorhynchus nerka}) in the Taku River, and Section~\ref{taku_discussion} concludes with a discussion and future directions.

\section{Methodology} \label{taku_method}

We now formalize the problem setup introduced in Section~\ref{taku_intro} and present our modeling framework. We observe estimated sample proportions $\mmu = (\hat{p}_1, \ldots, \muk)'$ with associated standard errors $\bs = (s_1, \ldots, s_K)'$ from a sample of size $n$ drawn from a population with true composition $\pp = (\pi_1, \ldots, \pi_K)'$ where $\sum_{k=1}^{K} \pi_k = 1$.

Let $\XX=(X_1,\ldots,X_K)'$ denote the unobserved sample counts for each group with $\XX\sim \mulnom(n, \pp)$, and let $\bm{\rho} = (\rho_{1}, \ldots, \rho_{K})' = \XX / n$ represent the corresponding sample proportions. We assume 
\begin{gather}
    \E(\mmu\mid\XX)=\bm{\rho}, \label{takueq:1stmom}\\
    \Var(\muk\mid X_k)=\sgk^2, \label{takueq:2ndmom}
\end{gather}
treating $\sgk$ as known, following common practice in meta-analysis of aggregate data~\citep{borenstein2021introduction}. Our goal is to derive the likelihood 
\begin{equation} \label{takueq:lik}
    p(\mmu\mid\pp) = \int p(\mmu,\XX\mid\pp)d\XX,
\end{equation}
which links the reported sample-level estimates $\mmu$ to the population parameter $\pp$ and forms the basis of our modeling approach. Since $\mmu$ is the output of an IA pipeline rather than a fully specified data-generating model, the exact form of $p(\mmu,\XX\mid\pp)$ is not available. We therefore approximate it with a generative model that satisfies the moment conditions in Equations~\eqref{takueq:1stmom} and~\eqref{takueq:2ndmom}, and yields a tractable expression for $p(\mmu\mid\pp)$.

\subsection{Reverse Dirichlet-Multinomial Model}\label{taku_amd}

Building on the setup described above, we now specify a probabilistic model for the distribution of $\mmu$ given $\pp$. The Dirichlet distribution is a standard choice on the simplex, so we propose a generative model for the compositional data as follows:
\begin{gather}
    \mmu\mid\XX \sim \dirch(\lambda\bm{\rho}), \label{takueq:simmodel}\\
    \XX \sim \mulnom(n, \pp),\label{takueq:simmodel2}
\end{gather}
where $\lambda>0$ is a concentration parameter controlling the variance of the Dirichlet distribution. We refer to this as the reverse Dirichlet-multinomial (RDM) model, with ``reverse'' indicating that the observed proportions, rather than parameters of interest, follow a Dirichlet distribution, distinguishing it from the conventional Dirichlet-multinomial model. 

Under the Dirichlet specification, dependence among the components of $\mmu$ given $\XX$ (equivalently, given $\bm{\rho}$) arises solely from the simplex constraint, with no additional source of correlation. This property is sometimes described as the ``ultimate in independence hypotheses''~\citep{aitchison1986}, in contrast to, for example, a multivariate logistic normal distribution. Under the RDM model, the conditional expectation and variance of $\muk$ given $\XX$ and $\lambda$ are:
\begin{align}
    \E(\muk\mid\XX,\lambda) &= \frac{\lambda\rho_k}{\sum_i \lambda\rho_{i}} = \frac{\lambda\rho_k}{\lambda} = \rho_k,\nonumber\\
    \Var(\muk\mid\XX,\lambda) &= \frac{\frac{\lambda\rho_k}{\sum_i \lambda\rho_{i}}(1-\frac{\lambda\rho_k}{\sum_i \lambda\rho_{i}})}{\sum_i \lambda\rho_{i}+1}= \beta\rho_k(1-\rho_k),\label{takueq:var}
\end{align}
where ${\beta = (\lambda+1)^{-1}}$. The expectation and variance of $\mmu$ given $\pp$ and $\lambda$ can be further derived:
\begin{align}
    \E(\mmu\mid\pp,\lambda) &=  \E\{\E(\mmu \mid \XX,\lambda)\} = \E(\XX/n) = \pp,\label{takueq:pimean}\\
    \Var(\mmu\mid\pp,\lambda) &= \E\{\Var(\mmu \mid \XX,\lambda)\} + \Var\{\E(\mmu \mid \XX,\lambda)\}\nonumber\\
    &=\E\left[\beta\{\text{diag}(\XX)/n - \XX\XX'/n^2\}\right] + \Var(\XX/n)\nonumber\\
    &=\beta(1-1/n)\left\{\text{diag}(\pp) - \pp\pp'\right\} + \frac{\text{diag}(\pp) - \pp\pp'}{n}\nonumber\\
    &= \tilde{\beta}\left\{\text{diag}(\pp) - \pp\pp'\right\},\label{takueq:pivar}
\end{align}
where
\begin{equation}
    \tilde{\beta}=\frac{n-1}{n}\beta + \frac{1}{n},\label{takueq:betatilde}
\end{equation}
is a decreasing function of $n$, bounded between $\beta$ and 1.

\subsection{Selecting $\lambda$ for the RDM Model}\label{taku_lambda}

With the RDM model, the likelihood in Equation~\eqref{takueq:lik} becomes $p(\mmu\mid\pp,\lambda)$. One may jointly estimate $(\pp,\lambda)$ from this likelihood. Here we instead adopt a plug-in estimate of $\lambda$ calibrated from the empirical mean-variance relationship between $\muk$ and $\sgk^2$, which is straightforward to compute and aligns with aggregate IA outputs.

In practice, a single value of $\lambda$ cannot, in general, reproduce the reported standard errors $\bs$ for all groups~\citep{aitchison1986,gelman1995method}. To address this, we propose selecting $\lambda$ by minimizing the discrepancy between the variance implied by the Dirichlet distribution and the observed variances in the data. This approach leverages the Dirichlet mean-variance relationship: $\Var(\muk\mid\XX,\lambda) \propto \E(\muk\mid\XX,\lambda)(1-\E(\muk\mid\XX,\lambda))$, as shown in Equation~\eqref{takueq:var}. We expect this relationship to approximately hold on the data as well, i.e., $s^2_k \approx {\beta\muk(1-\muk)}$, indicating that estimated variances $s^2_k$ have an approximate quadratic relationship with $\muk$ and approach zero when $\muk$ is near 0 or 1. This pattern can be assessed by plotting $s^2_k$ against $\muk(1-\muk)$. In our application in Section~\ref{taku_application}, this pattern is closely satisfied (Figure~\ref{takufig:linearity}). However, if the assumption is violated, the Dirichlet distribution may not be appropriate for these data, and proceeding with it may not be advisable.

Formally, we estimate $\beta$ via Equation~\eqref{takueq:lambda}, by minimizing the squared deviations between the observed variances in the GSI dataset and those implied by the Dirichlet model, where $\muobsk$ and ${\sgobsk}^2$ denote the observed stock proportions and corresponding variances:
\begin{equation}
    \hat{\beta} = \arg\min_{\beta}\sum_{k=1}^K\left\{{\sgobsk}^2-\beta\muobsk(1-\muobsk)\right\}^2. \label{takueq:lambda}
\end{equation}

The solution to Equation~\eqref{takueq:lambda} corresponds to the least squares estimator of $\beta$ in a simple linear regression model without intercept, ${\sgobsk}^2 = \beta\muobsk(1-\muobsk)+\epsilon_{k}$, $k=1,\dots,K$, where $\muobsk(1-\muobsk)$ is the predictor and $\epsilon_{k}$ is an error term. When $\hat{\beta}$ is used in inference, $\beta$ in Equation~\eqref{takueq:betatilde} is substituted by $\hat{\beta}$ and $\tilde{\beta}$ is bounded by $\hat{\beta}$ and 1. When $n$ is large, $\bs$ tends to be small, driving $\hat{\beta}$ towards 0 and hence $\tilde{\beta}$ towards 0. Hereafter, we consider $\tilde{\beta}$ as a function of $\hat{\beta}$, instead of $\beta$, unless otherwise specified. 

Finally, with $\beta=(\lambda+1)^{-1}$, the plug-in concentration parameter is
\begin{equation}\label{takueq:linearity}
    \hat{\lambda} = \hat{\beta}^{-1}- 1= \frac{\sum_k\{\muobsk(1-\muobsk)\}^2}{\sum_k \muobsk(1-\muobsk)(\sgobsk)^2}-1.
\end{equation}

\subsection{Moment-Matching Dirichlet Model and Variance Estimation}\label{taku_dirichlet}

As described in Section~\ref{taku_amd}, Equations~\eqref{takueq:pimean} and~\eqref{takueq:pivar} provide the mean and variance of $\mmu$ given $\pp$. These coincide with those of the Dirichlet model 
$$\mmu\sim\dirch(\tilde{\lambda}\pp), $$
where $\tilde{\lambda} = \tilde{\beta}^{-1}-1$. This observation motivates a computationally simpler approximation to the RDM model, which avoids integrating over $\XX$. We refer to this model as the moment-matching Dirichlet (MMD) model. 

Quantifying uncertainty under $p(\mmu\mid\pp,\lambda)$ hinges on the sampling variance of each component $\sigma^2_k \equiv \Var(\muk\mid\pp)$. Both the RDM and MMD models imply $\sigma^2_k=\tilde{\beta}\pi_k(1-\pi_k)$, so a direct plug-in estimator is $\hat{\sigma}^2_k = \tilde{\beta}\muk (1-\muk)$. An alternative estimator, which leverages the approximate proportional relationship between $s_k^2$ and $\muk (1-\muk)$ described in Section~\ref{taku_lambda}, is $\sgalt = \frac{\tilde{\beta}}{\hat{\beta}} s_k^2 = \left(1+\frac{\lambda}{n}\right) s_k^2 $. While it has the desirable property $\sgalt \geq s_k^2$, this property is not guaranteed to hold for $\hat{\sigma}^2_k$. Hence, $\sgalt$ may be more robust to deviations from the quadratic mean-variance relationship, as it directly incorporates $s_k^2$ and is less sensitive to inaccuracies in $\muk$. Another significant characteristic of $\sgalt$ is that for large $n$, $\sgalt \approx s_k^2$. Finally, it is crucial to emphasize that both variance estimators rely on the assumption that the simplex constraint is the sole source of covariance between groups when assuming a Dirichlet distribution.

\section{Methods for Genetic Mark-Recapture Studies} \label{taku_gmr}

We now extend our framework to genetic mark-recapture (GMR) studies, where the primary goal is to estimate the total escapement $N$ over the migration season in mixed-stock fisheries, i.e., the total number of fish returning to spawn across all stocks. Because GMR studies typically span multiple weeks, it is often useful to express escapement at finer resolutions. For example, the escapement for a specific group $k$ during week $t$ can be written as $N\w\ptk$, where $\w$ represents the proportion of the run (i.e., the total seasonal migration) occurring in week $t$, known as the ``run weight'', and $\ptk$ is the proportion of the run attributable to group $k$ in that week. Aggregates for specific weeks or groups can be obtained by summing $N\w\ptk$ over the relevant $t$ and/or $k$. 

Since compositional GSI data are collected weekly, all notations from Section~\ref{taku_method} now includes a week index $t \in {1,\ldots,T}$. Let $M = \sum_{t=1}^T\sum_{k=1}^{L} N\w\ptk$ denote the total escapement of lake-type stocks, where $L$ represents the number of lake-type stocks. As in usual practice, we treat $M$ and $\w$ as known without error, based on auxiliary surveys~\citep{gazey}.

\subsection{Existing Method}\label{taku_gazey}

Before introducing our proposed approach, we first review a commonly used estimator in GMR studies---the method-of-moments estimator proposed by~\citet{gazey}:
$$
\Nhat= \frac{M}{\sum_t\w\mut^\text{lake}},
$$
where the denominator estimates the overall proportion of lake-type stock in the run, with $\mut^\text{lake} = \sum_{k=1}^L \hat{p}_{k,t}$. 

The conditional variance of $\Nhat$ can be approximated using a first-order Taylor series expansion as 
$$
\Var(\Nhat | \bw)\approx\frac{M^2}{\{\sum_t\w\E(\mut^\text{lake}| \bw)\}^4}\sum_t\w^2\Var\left(\mut^\text{lake}| \bw\right).
$$
\citet{gazey} further proposed a variance estimator
\begin{equation*}
    \widehat\Var(\Nhat) = \left(\frac{\Nhat}{\sum_t\w\mut^\text{lake}}\right)^2\sum_t \w^2(s_t^\text{lake})^2,
\end{equation*}
where $\E(\mut^\text{lake}| \bw)$ is approximated by $\mut^\text{lake}$ and $\Var(\mut^\text{lake}| \bw)$ is approximated by the sample-level variance $(s_t^\text{lake})^2$. Here, $\mut^\text{lake}$ is an estimate for $\pi_t^\text{lake} = \sum_{k=1}^{L} \ptk$, and $s_t^\text{lake}$ denotes the sample-level standard error in estimating the lake-type proportion.  

However, it is important to point out that the population-level variance, $\Var(\mut^\text{lake}| \bw)$ is $\E(\Var(\mut^\text{lake}| \bw, \bm{g})| \bw) + \Var(\E(\mut^\text{lake}| \bw, \bm{g})| \bw)$, which is larger than the sample-level variance, $\E(\Var(\mut^\text{lake}| \bw, \bm{g})| \bw)$, estimated with $(s_t^\text{lake})^2$, where $\bm{g}$ represents all the genetic information from the genetic samples. As a result, this variance estimator will underestimate the true variance, potentially leading to incorrect results.

\subsection{Bayesian Approaches}\label{taku_bayes}

Bayesian inference provides a flexible alternative to address these limitations. The hierarchical structure of both RDM and MMD models naturally facilitate Bayesian analysis. After obtaining posterior draws of $\ptk$, we compute the posterior distribution of the total escapement $N$ for the season by sampling from 
\begin{equation} \label{takueq:N}
    N=\frac{M}{\sum_{t=1}^{T}\w\sum_{k=1}^L\ptk}
\end{equation}
at each iteration of a using Markov Chain Monte Carlo (MCMC) sampler. We then take the posterior mean (or median) of $N$ as the point estimate, and the posterior variance as the measure of uncertainty. 

This approach relies on the standard GMR assumptions, such as a closed population (no entries, deaths, or migration of fish between the GSI sample and the lake counts), equal sampling probabilities within each week, accurate genotyping of collected samples, and forward movement of fish (no repeated counting across weeks). In addition, we assume that lake-type stocks are sufficiently abundant so that the $\ptk$ values from our model can be interpreted as population proportions rather than super-population parameters.

\subsubsection{Prior Specification for \texorpdfstring{$\bm{\pi}$}{pi}} \label{taku_prior}

To complete the Bayesian formulation, we specify a prior distribution for $\pp$. A simple choice is the independent Dirichlet(\textbf{1}) prior within each week, which is diffuse on the simplex but may still exert undue influence when GSI sample sizes are small. To better reflect temporal dependence and share information across weeks, we introduce a time-series prior that captures the correlation in $\ptk$ over time. 

Our prior is inspired by~\citet{gelman1995method}'s transformed normal prior for parameters constrained to sum to one, originally applied in physiological pharmacokinetic models in~\citet{gelman1996physiological}. For a composition vector $(\pi_1,\ldots,\pi_K)$, each element is defined as $\pi_k=\frac{e^{Y_k}}{\sum_{i=1}^K e^{Y_i}}$, where each $Y_k$ follows an independent normal distribution, $Y_k\sim N(\theta_k,\psi_k^2)$. We propose a natural AR(1) extension of this formulation as follows:
\begin{align*}
    \ptk &= \frac{e^{Z_{k,t}}}{\sum_{i=1}^K e^{Z_{i,t}}}, \text{ for }t=1,\ldots,T, k=1,\ldots,K\\
    Z_{k,t} &=\phi Z_{k,t-1}+\epsilon_{k,t}, \text{ for }t=2,\ldots,T, k=1,\ldots,K\\
    \epsilon_{k,t}&\iid N(0,(1-\phi^2)\psi^2), \text{ for }t=2,\ldots,T, k=1,\ldots,K,
\end{align*}
with hyperpriors $Z_{k,1}\iid N(0,\psi^2)$ and $\phi\sim\text{Unif}(-1,1)$. Determining an appropriate value for $\psi$ is crucial to ensure that $\ptk$ can encompass values across the entire range of $[0,1]$. Details on this calibration are provided in Appendix~\ref{taku_arprior}. 

\subsection{Frequentist Approaches} \label{taku_freq}

To mitigate the variance underestimation in~\citet{gazey}, we substituted $(s_t^\text{lake})^2$ with $(\hat{\sigma}_t^\text{lake})^2 = \betat\hat{p}_t^\text{lake}(1-\hat{p}_{t}^\text{lake})$. This estimator represents a clear advancement compared to the one used in~\citet{pscreport}, where $\tilde{\beta}_t$ was defined as $\frac{1}{n^\text{eff}_t}$, with effective sample size $n^\text{eff}_t = \n(a + b \mut^\text{lake})$ for some estimated constants $a$ and $b$. In their formulation, as the sample size $\n$ grows, $(\hat{\sigma}_t^\text{lake})^2$ decreases towards 0, which is not desirable since the variance should converge to the population-level variance rather than vanish. 

An alternative is to use $(\hat{\sigma}_t^\text{lake, alt})^2 = \frac{\betat}{\hat{\beta}_t}(s_t^\text{lake})^2 = \left(1+\frac{\lam}{\n}\right)(s_t^\text{lake})^2$, where the factor $1+\lam/\n$ serves as an inflation term for the sample variance. This expression follows from the mean-variance relationship of the Dirichlet distribution described in Section~\ref{taku_dirichlet}.

In practice, the standard error $s_t^\text{lake}$ is often not reported in published data. In such case, we estimate it based on our earlier assumption that the correlation between regions is solely due to the simplex. Specifically, we calculate a pooled standard error for lake-type proportion as 
$$
s_t^\text{lake} = \max\left(\sqrt{\sum_{k=1}^L s_{k,t}^2 - 2\sum_{k=1}^L\sum_{l={k+1}}^L\betat\hat p_{k,t}\hat p_{l,t}}, \sqrt{\sum_{k=1}^L s_{k,t}^2 - 2\sum_{k=1}^L\sum_{l={k+1}}^Ls_{k,t}s_{l,t}}\right),
$$
where the first term is the variance estimate implied by the Dirichlet assumption, and the second argument is the theoretical lower bound corresponding to a correlation of -1 between stocks.

\section{Simulation Study} \label{taku_sim}

To assess the performance of our methods, we conducted a simulation study designed to closely mimic a real-world GMR scenario. Parameter values were selected to resemble the Taku River Sockeye Salmon data discussed in Section~\ref{taku_application}. Specifically, we used the observed weekly stock proportions $\muobs$ and standard errors $\sgobs$ from the Taku River GSI data to compute $\lam$ via Equation~\eqref{takueq:linearity}. The same $\muobs$ values were taken as the true values of $\ptk$ in simulations. We set the true total escapement $N$ to 60,000 (close to the estimate in~\citet{pscreport}), and calculated the lake-type escapement count $M=41,326$ using Equation~\eqref{takueq:N}. The migration season was assumed to span $T=12$ week, with $K=4$ distinct stocks, of which $L=2$ were lake-type and the remaining two were river-type, similar to the regional composition of the Taku River dataset. 

Simulated datasets were generated under the RDM model by first drawing $\bm{\rho}_t$ from the multinomial distribution (Equation~\eqref{takueq:simmodel2}) and then $\mmu_t$ from the Dirichlet distribution (Equation~\eqref{takueq:simmodel}), with corresponding standard errors $\sgtk$. A challenge during simulation was the potential for simulated $\mutk$ values to reach extremes like 0 or 1, which could lead to errors in downstream calculations. Additionally, if one regional stock dominates certain weeks in the GSI dataset, $X_{k,t}$ could be 0, causing parameters in the Dirichlet model to be 0 and resulting in errors. To mitigate these issues, we implemented two constraints during the simulation process using rejection sampling, retaining only the simulated datasets that met the conditions: $\mutk\in[10^{-10},1-10^{-7}]$ and $\rhotk>10^{-10}$. These thresholds were chosen to prevent degenerate variance calculations while minimally restricting the parameter space.

Both Bayesian and frequentist implementations of the RDM and MMD models were evaluated. In Bayesian approaches, we investigated the impact of two types of priors: a diffuse Dirichlet$(\bm{1})$ prior and the AR(1) prior described in Section~\ref{taku_prior}. For each simulated dataset, we recalculated $\lam$ to derive $\lamt$ from the simulated $\mutk$ and $\sgtk$. Posterior samples of the total escapement $N$ were obtained at each MCMC iteration using Equation~\eqref{takueq:N}, based on the posterior draws of $\ptk$. For the frequentist methods, we applied three variance estimators: $(\hat{\sigma}_t^\text{lake})^2$ calculated as $\betat\hat{p}_t^\text{lake}(1-\hat{p}_{t}^\text{lake})$, $(\hat{\sigma}_t^\text{lake, alt})^2$, and $(s_t^\text{lake})^2$. The last one is a naive variance estimator that does not propagate uncertainty from the sample level to the population level. 

For each simulation study setting, we generated and analyzed 1,000 datasets using R~\citep[Version 4.3.1;][]{R}, using a fixed seed for reproducibility. Bayesian methods were implemented using JAGS~\citep[Version 4.3.2;][]{plummer2003jags}, where we ran three chains until the second half of the chains showed convergence, confirmed by a Gelman-Rubin convergence diagnostic $\hat{R}<1.1$. Furthermore, the chains were thinned to 10,000 iterations to reduce storage space. 

Performance metrics for the total escapement $N$ included the relative Monte Carlo bias of the estimates (RBias), the relative root mean squared error of the estimates (RRMSE), as well as the average length (LCI), and coverage probability (CP) of the 95\% quantile-based credible interval for Bayesian methods, or $\Nhat\pm 1.96\cdot\sqrt{\widehat\Var(\Nhat)}$ for the method-of-moments in frequentist approaches.

\subsection{Results} \label{taku_res}

Table~\ref{takutab:res} presents a comprehensive overview of the estimated escapement across the simulation settings. The top four rows correspond to the proposed RDM and MMD models under two prior specifications, while the bottom three rows show results from the method-of-moments (MoM) estimator with alternative variance estimators discussed in Section~\ref{taku_freq}. Average computation times per dataset are also reported to assess computational efficiency.

\begin{table}[!htbp]
    \centering
    \caption{Comparison of estimation methods using results from the simulation study. ``RDM'' denotes the reverse Dirichlet-multinomial model, ``MMD'' represents the moment-matching Dirichlet model,  ``MoM'' refers to the method-of-moments estimator with $(\hat{\sigma}_t^\text{lake})^2=\betat\hat{p}_t^\text{lake}(1-\hat{p}_{t}^\text{lake})$, ``MoM(Alt)'' refers to the method-of-moments estimator using $(\hat{\sigma}_t^\text{lake, alt})^2$, and ``MoM(Naive)'' refers to the method-of-moments estimator using $(s_t^\text{lake})^2$ instead of $(\hat{\sigma}_t^\text{lake})^2$. \vspace{3pt}}
    \label{takutab:res}
    \begin{tabular}{cccccccc}
        \hline
        Model & Prior & RBias & RRMSE & CP & LCI & Time/Dataset\\
        \hline
        RDM & Dir & 0.026 & 0.039 & 0.88 & 7121 & 33.7 secs\\
        RDM & AR(1) & 0.010 & 0.032 & 0.95 & 7182 & 44.6 secs\\
        MMD & Dir & 0.015 & 0.032 & 0.95 & 7056 & 0.3 secs\\
        MMD & AR(1) & 0.005 & 0.031 & 0.95 & 7172 & 2.9 secs\\
        MoM & N/A  & -0.001 & 0.030 & 0.94 & 6987 & 0.001 secs\\
        MoM(Alt) & N/A & -0.001 & 0.030 & 0.96 & 8046 & 0.001 secs\\
        MoM(Naive) & N/A & -0.001 & 0.030 & 0.76 & 4174 & 0.002 secs\\
        \hline
    \end{tabular}
\end{table}

The results demonstrate that the choice of prior significantly influences the performance of the proposed models. Specifically, AR(1) priors consistently reduce relative bias and RRMSE, along with coverage probabilities closer to the nominal 95\%, compared to the Dirichlet prior. This improvement can be attributed to the weekly regional proportions in the Taku dataset used in our simulation study, as shown in Figure~\ref{takufig:regionmean}. The regional proportions vary significantly across weeks and exhibit temporal trends in at least three regions. In such scenarios, a diffuse Dirichlet prior may not adequately capture the underlying dynamics, hence affecting the model's performance.

Additionally, the comparison between different models for Bayesian inference indicates that the MMD model with AR(1) prior demonstrates superior performance in terms of RBias, RRMSE, and CP. Moreover, the MMD model significantly reduces computation time, achieving over a 15-fold reduction compared to the RDM model with the AR(1) prior, and approximately a 100-fold reduction compared to the same model with the Dirichlet prior, while maintaining comparable statistical performance.

The frequentist MoM estimators exhibit the smallest relative bias, consistent with their unbiasedness in expectation, and have RRMSE values similar to the Bayesian methods. However, the MoM estimator with the naive variance estimator (MoM(Naive)) has substantially lower coverage (76\%), highlighting the inadequacy of directly using the sample-level variance as the population-level variance. This finding emphasizes the importance of correctly estimating variance to achieve reliable coverage probabilities, as discussed in Sections~\ref{taku_intro} and~\ref{taku_gmr}. Furthermore, the average lengths of the 95\% credible or confidence intervals from the Bayesian methods fall between those from MoM and MoM(Alt), suggesting that the Bayesian methods provide a balanced estimation of uncertainty compared to the frequentist estimators, although the differences are subtle and may be attributed to Monte Carlo error.

We further explore the ability of our Bayesian models to estimate the proportions $\ptk$. Figures~\ref{takufig:mmdarp}, and~\ref{takufig:mmddirp},~\ref{takufig:amdarp},~\ref{takufig:amddirp} in Appendix~\ref{taku_supfig} present results for MMD with AR(1) prior, MMD with Dirichlet prior, RDM with AR(1) prior, and RDM with Dirichlet prior, respectively. Model MMD generally performs better than RDM and the AR(1) prior outperforms the Dirichlet prior, which is consistent with the results above.

\begin{figure}[!htb]
    \centerline{
    \includegraphics[width=6in]{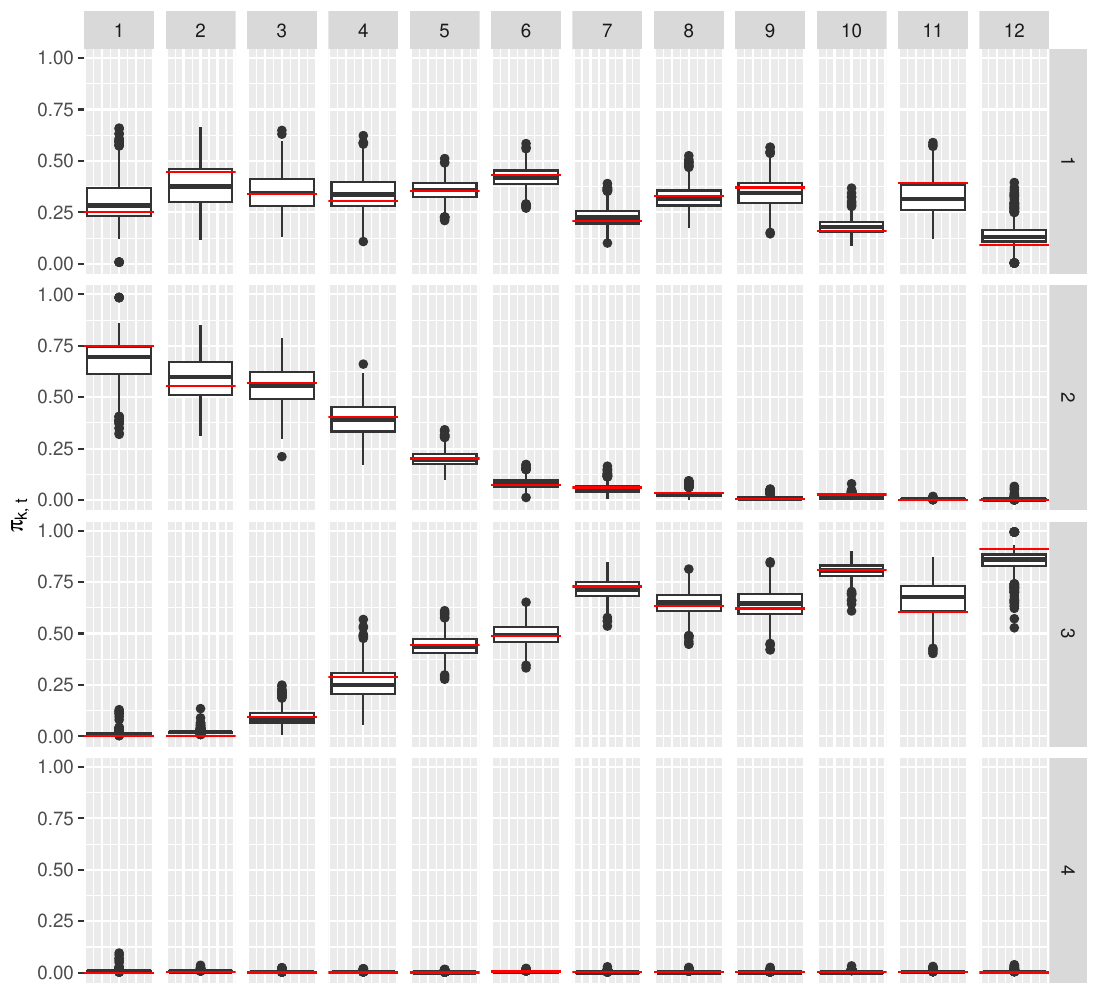}}
    \caption{Distribution of the posterior mean for $\ptk$ from the MMD model with AR(1) prior, based on 1,000 simulated datasets. Each panel corresponds to a specific stock $k=1,\ldots,4$ (rows) and week $t=1,\ldots,12$ (columns). Red horizontal lines indicate the true values of $\ptk$, used in the simulation.}
    \label{takufig:mmdarp}
\end{figure}

\section{Application to the Taku River Salmon Run Estimation} \label{taku_application}

\begin{figure}[!htb]
    \centerline{
    \includegraphics[width=\textwidth]{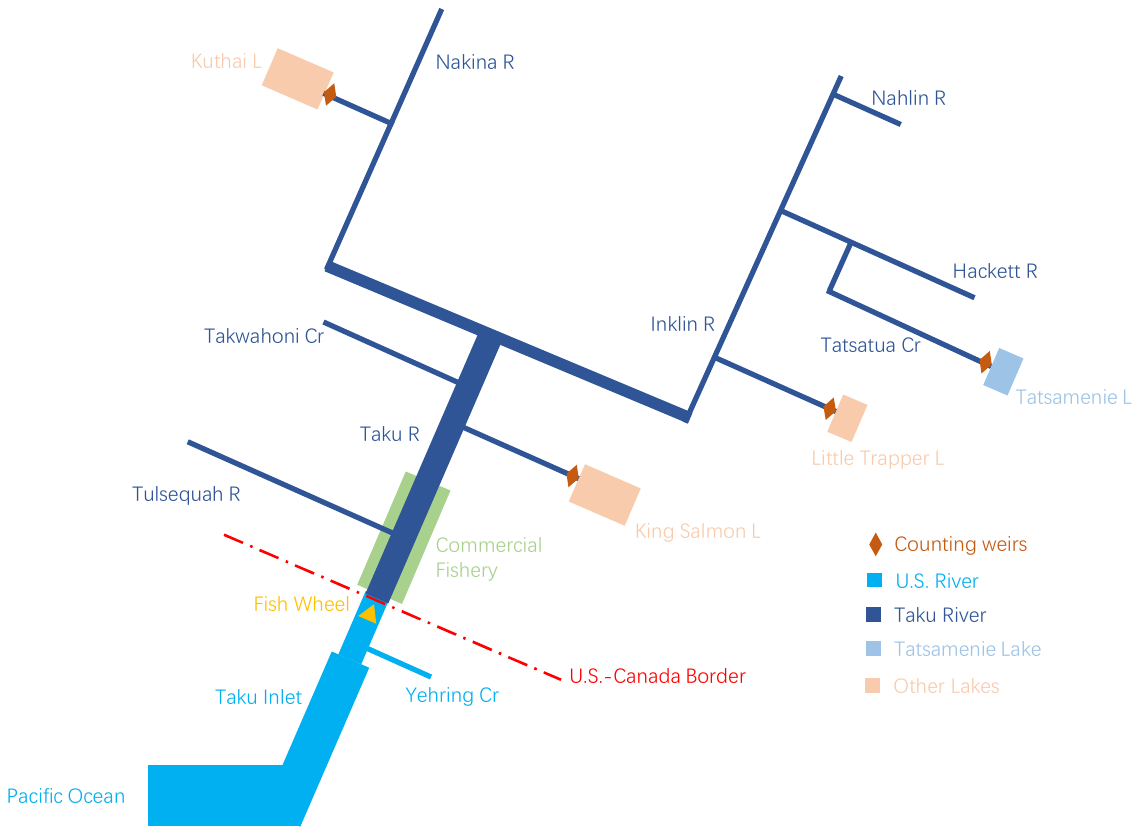}}
    \caption{Overview of Taku River system and study settings. R denotes river, Cr denotes creek and L denotes lake.}
    \label{takufig:overview}
\end{figure}

The Taku River is a river system originating from the Stikine Plateau in northwestern British Columbia, Canada, flowing into Juneau, Alaska. Renowned for its productivity, the Taku River hosts one of the largest runs of Sockeye Salmon in Southeast Alaska and northern British Columbia. Figure~\ref{takufig:overview} provides an overview map of the Taku River system and the study settings.

Genetic data on Sockeye Salmon in the Taku River has been collected since 2008 by the Canadian commercial fishery and annually analyzed by the Molecular Genetics Laboratory of Fisheries and Oceans Canada (DFO). Starting in 2012, the U.S. District 111 Commercial Fishery also began collecting genetic data annually, with analysis conducted by the Gene Conservation Laboratory of the Alaska Department of Fish and Game (ADF\&G). These collaborative efforts have led to the development of a comprehensive genetic baseline for Taku River Sockeye Salmon, comprising 17 unique genetic groups, including 13 river-type and 4 lake-type stocks. Given that many of these stocks have very small proportions, often close to zero, there is a high likelihood that some stocks may not be included in a given sample. To mitigate this issue and ensure more reliable estimates, these stocks are further grouped into four regions/reporting units based on their type and geographic location: U.S. River, Taku River, Tatsamenie Lake, and Other Lakes, as illustrated in Figure~\ref{takufig:overview}. 

\begin{figure}
    \centering
    \begin{subfigure}[b]{0.48\textwidth}
         \centering
        \includegraphics[width=3.5in]{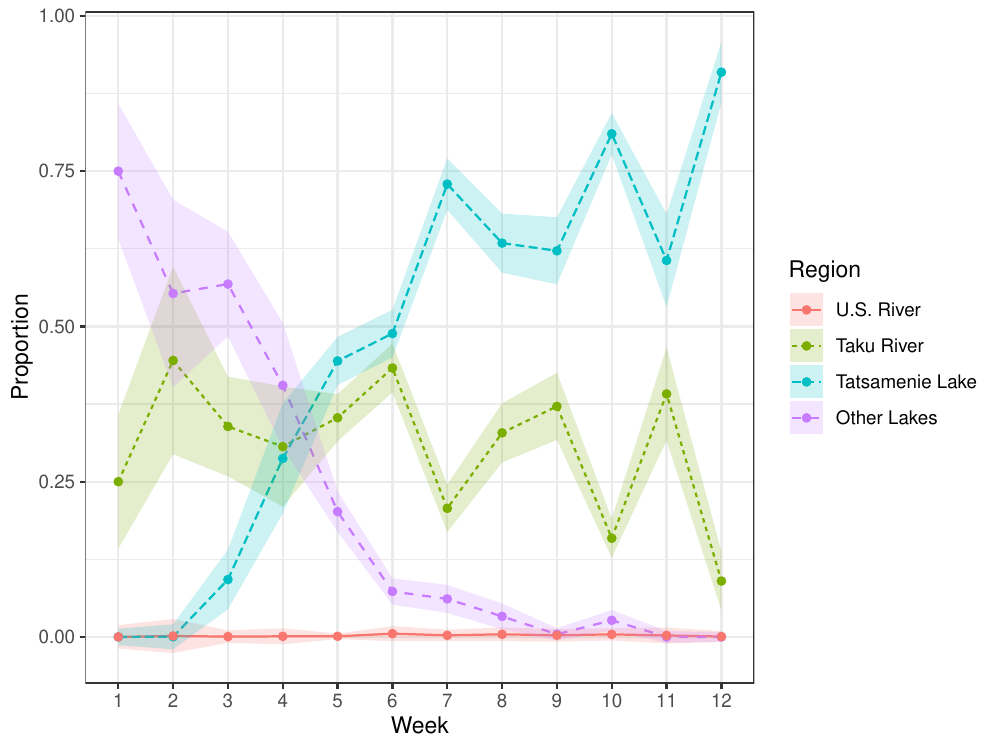}
        \caption{Weekly proportions ($\hat p_{k,t}^\text{obs}$) of genetic samples from each region $k$. Each dot in the scatter plot represents a mean proportion while shaded bands extend one standard error ($s^\text{obs}_{k,t}$) above and below the mean.}
        \label{takufig:regionmean}
    \end{subfigure}\hfill
    \begin{subfigure}[b]{0.45\textwidth}
        \centering
        \includegraphics[width=3.5in]{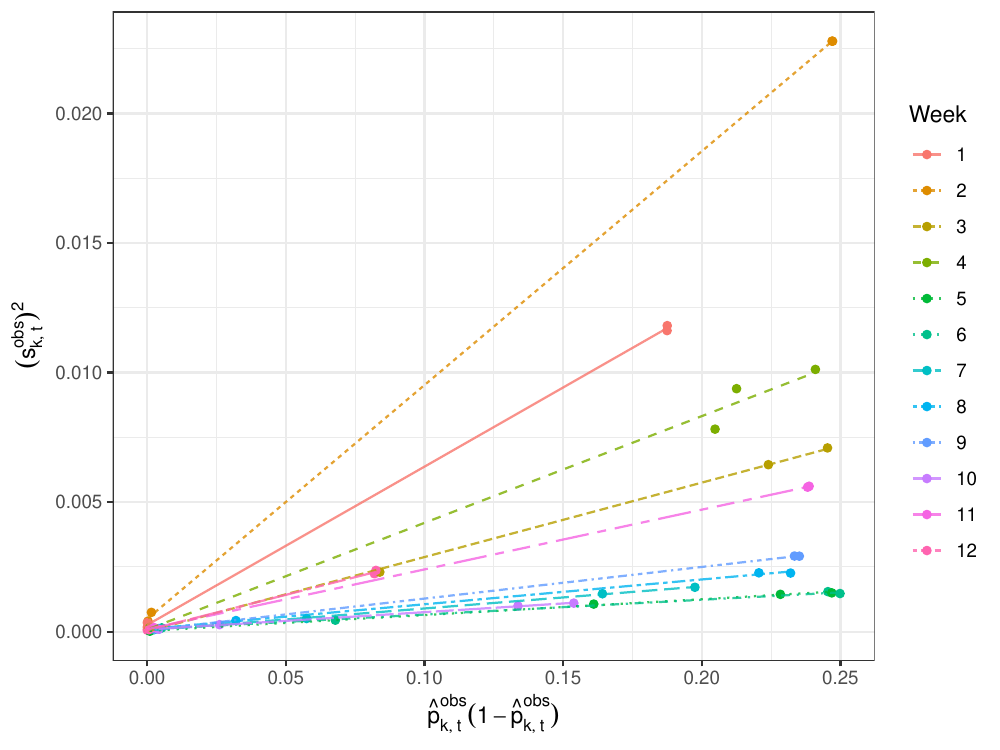}
        \caption{Weekly variances $(s^\text{obs}_{k,t})^2$ as a function of $\hat p_{k,t}^\text{obs}(1-\hat p_{k,t}^\text{obs})$, with regression lines for each week.\vspace{20pt}}
        \label{takufig:linearity}
    \end{subfigure}
\caption{Summary of observed GSI data in 2017.}
\end{figure}

We focus our analysis on the 2017 season as a representative case study. Table~\ref{takutab:weight} provides GSI sample sizes $\n$, while Figure~\ref{takufig:regionmean} presents GSI summary data from weekly samples at the regional level. These summaries were derived using the Bayesian GSI algorithm proposed by~\citet{Pella2001Bayesian}. The algorithm generates, for each fish in the sample, individual assignments (IA) to stocks via Markov Chain Monte Carlo. These IAs are then summarized by averaging over individuals to obtain composition estimates $\hat{p}_{k,t}$ along with their sample-level standard errors $s_{k,t}$. Figure~\ref{takufig:linearity} displays the relationship between $s^2_{k,t}$ and $\hat{p}_{k,t}(1-\hat{p}_{k,t})$ in each week $t=1,\ldots,12$. We observe that the weekly relationships exhibit a close proportionality, which is a crucial prerequisite for our Dirichlet assumption. Furthermore, we include the variance inflation factor $1+\lam/\n$, discussed in Section~\ref{taku_method} and Section~\ref{taku_freq}, in Table~\ref{takutab:weight}. We can see that the effect of this variance inflation factor remains relatively constant across weeks, reaching a maximum when the GSI sample size is 112 in week 7. However, if $\n$ were very large, we would expect this inflation factor to diminish towards 1.

\begin{table}[!htb]
    \centering
    \caption{Weekly weights ($\w$), GSI sample size ($\n$), and variance inflation factors ($1+\frac{\lam}{\n}$) for the 2017 Taku River Sockeye Salmon run.}
    \label{takutab:weight}
    \begin{tabular}{ccccccccccccc}
      \hline
    Week ($t$) & 1 & 2 & 3 & 4 & 5 & 6 & 7 & 8 & 9 & 10 & 11 & 12 \\ 
      \hline
      $\w$ & 0.02 & 0.01 & 0.04 & 0.03 & 0.18 & 0.18 & 0.11 & 0.11 & 0.09 & 0.16 & 0.04 & 0.04 \\ 
      $\n$ & 17 & 13 & 38 & 26 & 172 & 178 & 112 & 105 & 84 & 146 & 43 & 42\\ \hline 
      $1+\frac{\lambda_t(\bs)}{\n}$ & 1.88 & 1.76 & 1.89 & 1.89 & 1.92 & 1.91 & 2.01 & 1.94 & 1.94 & 1.93 & 1.97 & 1.83\\\hline
    \end{tabular}
\end{table}

In addition to genetic data collection, fish wheels are set up close to the U.S.-Canada border, providing weekly weights displayed in Table~\ref{takutab:weight}. Finally, counting weirs are constructed at the entrances of the lakes within the river system. Considering the salmon's predisposition to return to their native lakes and rivers, these counting weirs can nearly perfectly tally the fish entering the lakes, providing an accurate total count $M=34,351$ for the lake-type component of the run.

We analyzed the 2017 Taku River Sockeye Salmon data in multiple ways, comparing all the approaches implemented in the simulation study. The results on total escapement estimation are presented in Table~\ref{takutab:app}. Although the total escapement estimates are similar across models, several noteworthy differences emerge. Firstly, the posterior mean estimates with the AR(1) prior are closer to the method-of-moments estimate compared to those with Dirichlet prior. Secondly, the proposed models using the Dirichlet prior yield a smaller posterior standard deviation compared to those with the AR(1) prior, while the frequentist method with the naive variance estimator has the smallest standard deviation. Moreover, compared to the naive variance estimator, both variance estimators proposed in Section~\ref{taku_freq} provide a standard deviation closer to the posterior standard deviations of the Bayesian approaches.

\begin{table}[!htbp]
    \centering
    \caption{Escapement estimation results for the 2017 Taku River Sockeye Salmon run using Bayesian and frequentist methods. ``Estimate'' for the Bayesian model refers to the posterior mean of $N$. ``SD'' denotes the posterior standard deviation for the Bayesian methods, and the estimated standard deviation for frequentist methods with different variance estimators. The 95\% CI shows the 95\% quantile-based credible intervals for the Bayesian approaches and $\Nhat\pm 1.96\cdot\sqrt{\widehat\Var(\Nhat| \bw)}$ for the frequentist methods. ``Time'' represents the computing time for each setting.}
    \label{takutab:app}
    \begin{tabular}{ccccccc}
    \hline
    Model & Prior on $\bm \pi$ & Estimate & SD & 95\% CI & Time \\
    \hline
    RDM & Dirichlet & 51,164 & 1,480 & (48,439, 54,265) & 97 secs\\
    RDM & AR(1) & 50,367 & 1,482 & (47,655, 53,455) & 427 secs\\
    MMD & Dirichlet & 50,956 & 1,526 & (48,147, 54,116) & 21 secs\\
    MMD & AR(1) & 50,354 & 1,551 & (47,490, 53,536) & 104 secs\\
    MoM & N/A & 49,873 & 1,498 & (46,937, 52,809) & 0.02 secs\\
    MoM(Alt) & N/A & 49,873 & 1,685 & (46,570, 53,175) & 0.03 secs \\
    MoM(Naive) & N/A & 49,873 & 876 & (48,156, 51,589) & 0.05 secs\\
    \hline
    \end{tabular}
\end{table}

\section{Discussion} \label{taku_discussion}

In this study, we proposed a novel methodology to estimate population-level compositions from sample-level data, introducing the reverse Dirichlet-multinomial model and a moment-matching Dirichlet approximation to improve computational efficiency. These methods were extended to genetic mark-recapture studies for estimating total escapement in mixed-stock fisheries. Through a comprehensive simulation study and a real-data analysis based on the Taku River Sockeye Salmon fishery data, we evaluated our approaches and compared with the standard method-of-moments estimator.

While our Bayesian models exhibited reliable performance, yielding accurate estimates of escapement and providing insights into the temporal variation in stock composition, they come with limitations. First, the reverse Dirichlet-multinomial model incurs non-negligible computational cost, especially for large datasets. Second, the Dirichlet assumption of low dependence across groups may be restrictive in contexts where stocks are genetically similar or strongly negatively correlated, potentially leading to underestimation of uncertainty.

Despite these caveats, our framework has broad applicability in compositional inference scenarios where only point estimates and associated uncertainties are available. Future work may focus on relaxing the independence assumption by incorporating models that permit structured covariance, or on extending these ideas to settings with multiple sources of uncertainty or hierarchical sampling designs.

\section*{Acknowledgements}
The authors thank Dr. Carl J. Schwarz and the CANSSI Collaborative Research Team on ``Addressing Spatial and Computational Issues in Integrated Analysis of Modern Ecological Data'' for enabling this project. This research was supported by NSERC Discovery Grants RGPIN-2020-04364 (ML) and RGPIN-2019-04404 (AB).\vspace*{-8pt}

\bibliographystyle{chicago}
\bibliography{ref} 

\appendix

\section{Determining $\psi$ in the AR(1) Prior Specification for $\ptk$} \label{taku_arprior}

\begin{figure}[!htbp]
    \centerline{
    \includegraphics[width=\linewidth]{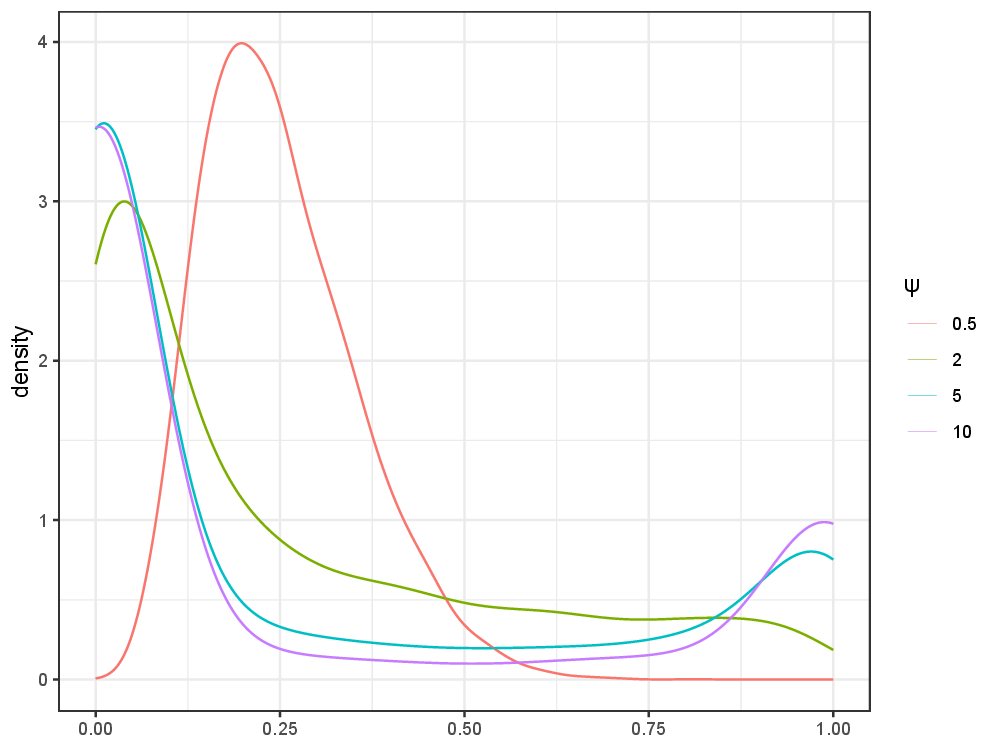}}
    \caption{Density plot for $\pi_{1,1}$ with different choices of $\psi$ in the time series prior.}
    \label{takufig:psi2}
\end{figure}

Determining an appropriate value for $\psi$ is crucial to ensure that $\ptk$ can encompass values across the entire range of $[0,1]$. To explore various options, we arbitrarily selected four values for $\psi$: 0.5, 2, 5, and 10, and conducted a simulation to assess their performance. With $K=4$ (as in the Taku River data example from Section~\ref{taku_application}), we simulated 10,000 $\pi_{1,1}$'s for each $\psi$ value. Figure~\ref{takufig:psi2} depicts the density plots for $\pi_{1,1}$ under different $\psi$ settings. It is evident that a small $\psi=0.5$ confines the range of $\pi_{1,1}$ mostly below 0.5, whereas $\psi \geq 5$ renders values between 0.25 and 0.75 quite improbable. Conversely, when $\psi=2$, the density curve is more evenly distributed across the range $[0,1]$, albeit showing an expected bias towards $1/K$. Consequently, $\psi=2$ emerges as a reasonable choice for the scenario $K=4$, which we choose for our analyses.

\section{Supplementary Figures}\label{taku_supfig}

\begin{figure}[!htb]
    \centerline{
    \includegraphics[width=6in]{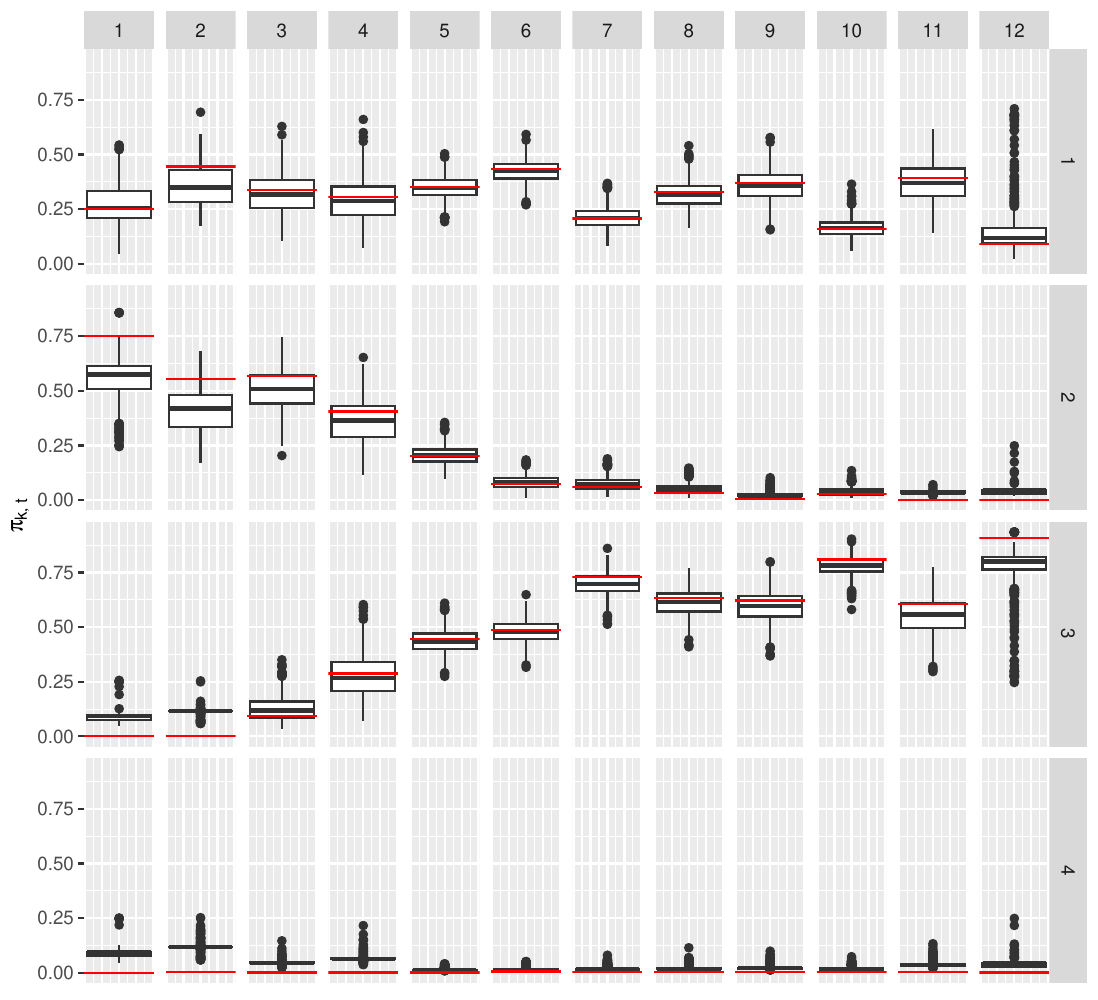}}
    \caption{Distribution of the posterior mean for $\ptk$ from the reverse Dirichlet-multinomial model with Dirichlet prior in the simulation study, for stocks $k=$ 1 to 4 in weeks $t=$ 1 to 12. Red horizontal lines indicate the true $\ptk$ values.}
    \label{takufig:amddirp}
\end{figure}

\begin{figure}[!htb]
    \centerline{
    \includegraphics[width=6in]{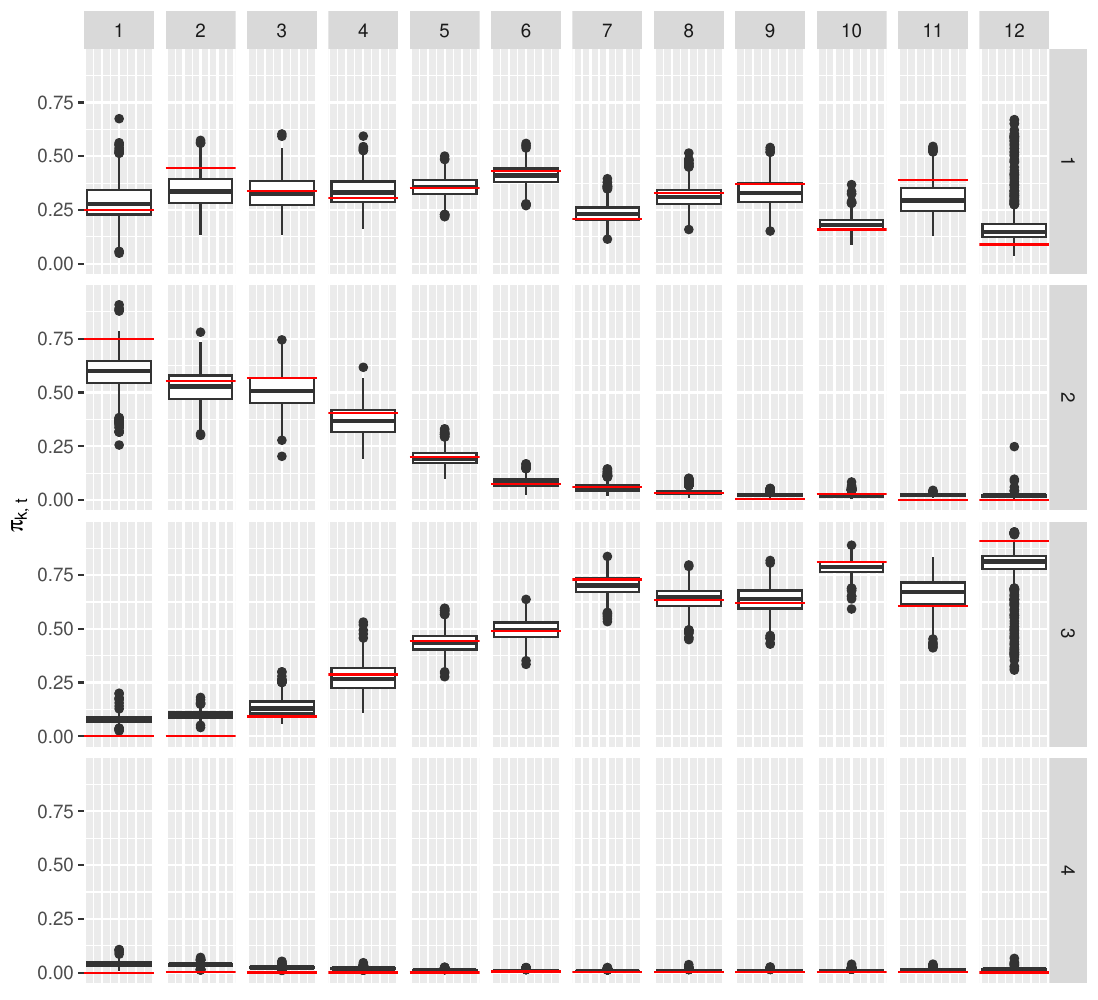}}
    \caption{Distribution of the posterior mean for $\ptk$ from the reverse Dirichlet-multinomial model with AR(1) prior in the simulation study, for stocks $k=$ 1 to 4 in weeks $t=$ 1 to 12. Red horizontal lines indicate the true $\ptk$ values.}
    \label{takufig:amdarp}
\end{figure}

\begin{figure}[!htb]
    \centerline{
    \includegraphics[width=6in]{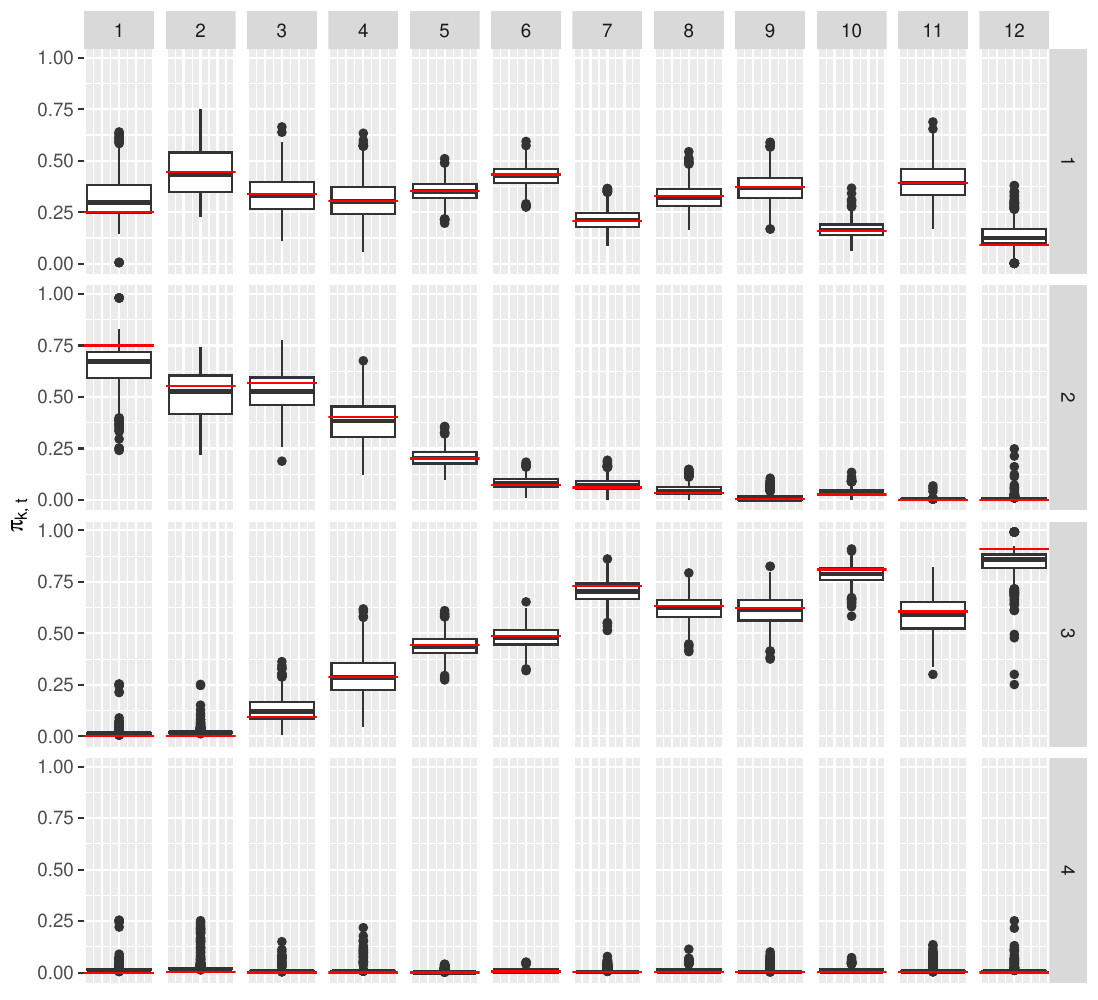}}
    \caption{Distribution of the posterior mean for $\ptk$ from the moment-matching Dirichlet model with Dirichlet prior in the simulation study, for stocks $k=$ 1 to 4 in weeks $t=$ 1 to 12. Red horizontal lines indicate the true $\ptk$ values.}
    \label{takufig:mmddirp}
\end{figure}

\end{document}